\documentclass[french]{gretsi}

\usepackage[utf8]{inputenc}

\usepackage[T1]{fontenc}

\usepackage{amsmath, amssymb, amsfonts}

\usepackage{comment}
\usepackage{utils}
\usepackage{multirow}

\usepackage{algorithmic}

\showboxdepth=\maxdimen
\showboxbreadth=\maxdimen

\begin{document}


\title{Modèle physique variationnel pour l'estimation de réponses impulsionnelles de salles}

\auteurs{
    \auteur{Louis}{Lalay}{louis.lalay@telecom-paris.fr}{}
    \auteur{Mathieu}{Fontaine}{mathieu.fontaine@telecom-paris.fr}{}
    \auteur{Roland}{Badeau}{roland.badeau@telecom-paris.fr}{}
}

\affils{
    \affil{}{Laboratoire Traitement et Communication de l'Information,\\
        Institut Polytechnique de Paris, 19 Place Marguerite Perey, 91120 Palaiseau, France
    }
}

\resume{
    Estimer la réponse impulsionnelle d'une salle est essentiel pour des tâches comme la déréverbération, qui améliore la reconnaissance automatique de la parole. La plupart des méthodes existantes reposent soit sur du traitement du signal statistique, soit sur des réseaux de neurones profonds s'inspirant du traitement du signal. Cependant, la combinaison des modélisations statistique et physique reste largement inexploré en estimation de réponse impulsionnelle de salle. Cet article propose une approche novatrice intégrant les deux aspects à travers un modèle physique. La réponse de salle est décomposée en paramètres interprétables : un bruit blanc gaussien modulé par une décroissance exponentielle dépendante de la fréquence (modélisant l'absorption des murs) et un filtre autorégressif (modélisant par exemple la réponse du microphone). L'optimisation d'une fonction d'énergie libre variationnelle permet une estimation pratique des paramètres. Nous montrons que, connaissant les signaux secs et réverbérants, la méthode proposée surpasse la déconvolution classique dans des environnements bruités, comme le confirment les mesures objectives.
}

\abstract{
    Room impulse response estimation is essential for tasks like speech dereverberation, which improves automatic speech recognition.
    Most existing methods rely on either statistical signal processing or deep neural networks designed to replicate signal processing principles.
    However, combining statistical and physical modeling for room impulse response estimation remains largely unexplored.
    This paper proposes a novel approach integrating both aspects through a theoretically grounded model. The room response is decomposed into interpretable parameters: white Gaussian noise modulated by a frequency-dependent exponential decay (e.g. modeling wall absorption) and an autoregressive filter (e.g. modeling microphone response). The optimization of a variational free-energy cost function enables practical parameter estimation.
    As a proof of concept, we show that given dry and reverberant speech signals, the proposed method outperforms classical deconvolution in noisy environments, as validated by objective metrics.
}

\maketitle


\section{Introduction}

L'estimation de réponse impulsionnelle de salle (RIR, de l'anglais \emph{Room Impulse Response}) est une tâche cruciale qui trouve des applications en adaptation acoustique~\cite{lee2023yet} et en reconnaissance automatique de la parole (ASR)~\cite{ratnarajah2023towards} par exemple. Une autre tâche clé liée à l'estimation de la RIR est la déréverbération, qui améliore les performances de l'ASR. Cependant, comme la déconvolution avec une RIR dégradée est très sensible, la plupart des méthodes de déréverbération évitent l'estimation explicite de la RIR et estiment directement le signal sec en utilisant des approches de traitement du signal~\cite{belhomme2017amplitude}, des approches stochastiques~\cite{nakatani2010speech, sekiguchi2022autoregressive}, ou des réseaux neuronaux profonds (DNN)~\cite{bahrman2024speech, hao2021fullsubnet}.

\begin{figure}[ht]
    \centering
    \includegraphics[width=0.75\linewidth]{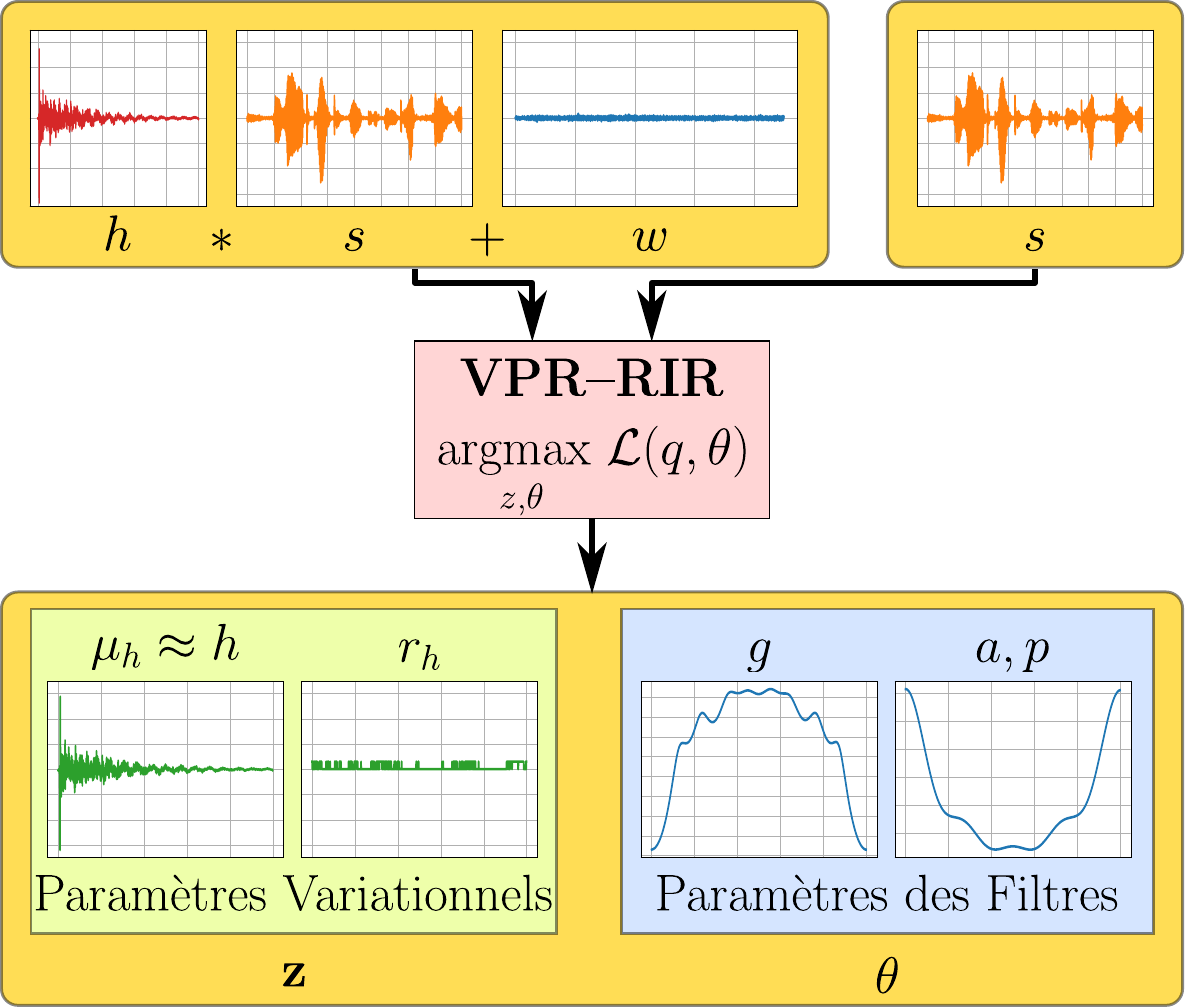}
    \caption{Aperçu du modèle physique variationnel de réverbération proposé pour l'estimation de RIR (\Our)}
    \label{fig:enter-label}
\end{figure}

Les techniques d'estimation de la RIR ont évolué au fil des décennies, introduisant divers modèles. L'un des premiers modèles a été proposé par Schröeder dans \cite{schroeder1962frequency}, où la RIR est représentée comme le produit d'un bruit blanc gaussien et d'une exponentielle décroissante. Plus tard, Polack \cite{polack1988transmission} a démontré que la composante gaussienne varie lentement dans le domaine fréquentiel, conduisant à une estimation plus précise de la RIR. Bien que ces modèles capturent efficacement la réverbération tardive, ils peinent à représenter avec précision les réflexions précoces.
Pour palier cette limitation, Komatsu \textit{et al.}~\cite{komatsu2013modeling} ont introduit un modèle de processus gaussien spatio-temporel qui prend en compte à la fois les réverbérations précoce et tardive. D'autres méthodes, comme celle de \cite{sundstrom2024estimation}, estiment la RIR en formulant des problèmes d'optimisation, en intégrant une régularisation par transport optimal.

Ces modèles n'estiment pas conjointement les informations spatiales, temporelles et fréquentielles, malgré leur corrélation théorique. Le travail de \cite{badeau2018unified} est une première tentative d'introduction d'un modèle spatio-temporel et dépendant de la fréquence. Sa première application a démontré qu'étant donnée une RIR réelle, il permet une estimation précise du temps de réverbération après une atténuation de 60dB ($\text{RT}_{60}$) \cite{aknin2021stochastic}.

Nous proposons dans cet article\footnote{Code accessible sur GitHub : \href{https://github.com/LouisLalay/VPR-RIR}{https://github.com/LouisLalay/VPR-RIR}} une formulation variationnelle utilisant le modèle physique de \cite{badeau2018unified} pour des RIR. Nous dérivons un critère d'énergie libre variationnelle qui sera utilisé pour estimer les paramètres du modèle. En particulier, la distribution variationnelle est supposée être une gaussienne dont la moyenne est la RIR cible. Dans cette première approche, nous supposons que les signaux secs et réverbérés sont connus, ce qui peut être vu comme une sous-tâche de l'adaptation acoustique. Le signal observé est corrompu par un bruit additif pour tester la robustesse de notre approche. Les résultats montrent la nette robustesse de la technique proposée par rapport à d'autres techniques classiques de déconvolution \cite{avargelSystemIdentificationShortTime2007}, selon divers critères caractéristiques de la réverbération ($\text{RT}_{60}$, \emph{energy decay relief}, \emph{energy decay curve}, etc.).
\section{Méthode}

Dans cette section, nous décrivons le modèle choisi, puis nous présentons le modèle de mélange considéré pour les expériences, et enfin l'approche variationnelle utilisée pour estimer les paramètres du modèle à partir des observations.

\subsection{Modèle de RIR et de mélange}

Soit $y \in \mathbb{R}^{T}$ le signal mesuré, $s \in \mathbb{R}^{L_s}$ le signal source et $h \in \mathbb{R}^{L_h}$ la réponse impulsionnelle de salle, de longueurs respectives $T, L_s$ et $L_h$, telles que $T = L_h + L_s - 1$. Nous considérons dans le domaine temporel le modèle de mélange suivant :

\begin{equation}
    y = s \ast h + w,
    \label{eq:mix}
\end{equation}
où $w \in \mathbb{R}^T$ est un bruit blanc gaussien additif de variance $\sigma_w^2$.
Nous modélisons le vecteur aléatoire $h$ comme dans \cite{aknin2021stochastic} :
\begin{equation}
    h = G^{-1}E^{-1}P^{-1}\epsilon
    \label{eq:rir}
\end{equation}
où les composantes sont définies comme suit :

\begin{itemize}
    \item $\epsilon[u] \sim\mathcal{N}(0,\sigma_\epsilon^2)$ pour $u = 0, \ldots, L_h-1$ (bruit blanc gaussien i.i.d.) ;
    \item $G = \text{Toep}(g)$, $g \in \mathbb{R}^{L_g}$, matrice triangulaire inférieure ;
    \item $E = \text{Diag}({\{e^{au}\}}_{u=0}^{L_h-1})$, $a\geq 0$, matrice diagonale ;
    \item $P = \mathcal{P}_{L_h}(p)$ avec $p_0=1$, $p \in \mathbb{R}^{L_p}$, matrice triangulaire inférieure.
\end{itemize}
Toutes les matrices sont carrées et de taille $L_h \times L_h$, correspondant à la longueur de la RIR en nombre d'échantillons. Toep($b$) et $\mathrm{Diag}(b)$ désignent respectivement la matrice de Toeplitz triangulaire inférieure et la matrice diagonale générée à partir d'un vecteur $b$.
$\mathcal{P}_{L_h}(p)$ est une matrice carrée de taille $L_h$ générée par $p$. La matrice $\mathcal{P}_4(p)$ est par exemple construite comme décrit ci-dessous, avec $L_h=4, p=[p_0,p_1]$ où $p^{*n} = p * p^{\ast(n-1)}$ avec $p^{\ast 0} = \delta$ (où $\delta$ désigne l'impulsion unitaire au temps $0$) et $p^{*n}[:l]$ le vecteur tronqué des $l$ premiers éléments de $p^{\ast n}$.

\[
    \begin{array}{c@{\quad}c}
                                      & \begin{array}{cccc}
                                            p^{\ast 0}[:4] & p^{\ast 1}[:4] & p^{\ast 2}[:4] & p^{\ast 3}[:4]
                                        \end{array} \\[10pt]

        \mathcal{P}_{4}([p_0, p_1]) = &
        \left( \begin{array}{c@{\qquad}c@{\qquad}c@{\qquad}c}
                       \mask{p^{*0}}{1} & \mask{p^{*0}}{0} & 0       & \mask{\hphantom{3p_0p_1^2}}{0} \\[4pt]
                       0                & p_0              & 0       & 0                              \\[4pt]
                       0                & p_1              & p_0^2   & 0                              \\[4pt]
                       0                & 0                & 2p_0p_1 & p_0^3
                   \end{array} \right).
    \end{array}
\]

Le modèle de RIR repose sur des hypothèses physiques et peut être décomposé en quatre parties. Tout d'abord, $\epsilon$ est le bruit blanc gaussien de base qui est filtré pour obtenir la RIR. Ensuite, la matrice de Toeplitz $G$, paramétrée par le vecteur $g$ de longueur $L_g < L_h$, implémente un filtre qui modélise la réponse du microphone. La matrice diagonale $E$ représente l'absorption moyenne de la salle, paramétrée par le coefficient d'absorption $a$. Enfin, la matrice $P$ représente l'absorption dépendante de la fréquence des murs, paramétrée par le filtre $p$ de longueur $L_p < L_h$. Avec trois matrices pour modéliser le filtre global, nous pouvons imposer $g_0=1$ et $p_0=1$ sans perte de généralité tout en conservant de bonnes propriétés en termes d'inversion et de calculs de déterminants. Le lien entre les propriétés physiques de la salle et les paramètres du modèle n'est pas clairement défini mais pourra être étudié dans des travaux futurs.

\subsection{Estimation des paramètres}

Nous adoptons une approche variationnelle pour estimer les paramètres du modèle. Soit $\theta = \{g,a,p,\sigma_\epsilon, \sigma_w\}$ l'ensemble des paramètres du modèle que nous cherchons à estimer.
La distribution cible, dans le cadre de la méthode d'approximation du champ moyen, pour $h$ que nous cherchons à approcher est donnée par $q^*(h[u]) = \mathcal{N}(\mu_{h}[u], r_{h}[u])$. Soit $z = \{ \mu_{h}, r_{h} \}$ l'ensemble des paramètres optimaux de la distribution, où $\mu_{h} , r_{h}$ sont les vecteurs contenant les composantes $\mu_h[u], r_h[u]$ respectivement. L'énergie libre variationnelle (VFE) associée à ce problème est définie comme suit \cite{bishop2006pattern} :

\begin{equation}
    \mathcal{L}(q,\theta) = \mathbb{E}_{q(z)}\left[\ln \frac{\mathbb{P}(y,z\mid\theta, s)}{q(z)}\right].\label{eq:VFE}
\end{equation}
En combinant le modèle stochastique de la RIR dans les Eqs.~\eqref{eq:rir} et \eqref{eq:mix}, nous obtenons une forme explicite pour l'Eq.~\eqref{eq:VFE} :

\begin{align}
    \mathcal{L}(q,\theta)
     & = - \frac{1}{2}\left(T\ln(2\pi\sigma_w^2)+\frac{\mathbb{E}_{q(z)} \left[\Vert y-s*h \Vert_2^2\right]}{\sigma_w^2}\right)\nonumber    \\
     & \hphantom{= {}}- \frac{1}{2}\left(L_h\ln(\sigma_\epsilon^{2})-L_h(L_h-1)a\right)         \nonumber                                   \\
     & \hphantom{= {}}- \frac{1}{2\sigma_\epsilon^{2}}\left(\Vert V\mu_h\Vert_2^2 + \mathrm{Tr}\left(V R_h {V}^\top\right)\right) \nonumber \\
     & \hphantom{= {}}+ \frac{1}{2}\left(\sum_{u=0}^{L_h - 1}\ln r_h[u] + L_h \right)
\end{align}
où $.^\top, \mathrm{Tr(.)}$ désignent respectivement la transposition et l'opérateur trace, $\Vert.\Vert_2$ représente la norme euclidienne d'un vecteur réel, $V=PEG$, $R_h= \text{Diag}(r_h)$, et
\begin{align*}
    \mathbb{E}_{q(z)} \left[||y-s*h||_2^2\right] = & ||y||_2^2 - 2 \sum_{t=0}^{T-1} y[t](\mu_h*s)[t] \\
    +                                              & \sum_{t=0}^{T-1} (r_h*s^2)[t] + (\mu_h*s)^2[t].
\end{align*}
Nous cherchons à maximiser cette fonction par rapport à $z$ et $\theta$. Cela peut être formulé comme un problème d'optimisation :

\begin{equation}
    (z, \theta) = \underset{z, \theta}{\text{argmax }} \mathcal{L}(q,\theta). \label{eq:optim_problem}
\end{equation}
Nous procédons ensuite à l'estimation des paramètres par descente de gradient sur la fonction $-2\mathcal{L}$. Dans les expériences, nous utilisons l'optimisateur Adam \cite{kingma2014adam} pour résoudre ce problème. Notez qu'un algorithme variationnel d'espérance-maximisation \cite{bishop2006pattern} peut également être adapté pour estimer les paramètres, ce que nous laissons pour des travaux futurs.

\subsection{Contrainte de cohérence sur \texorpdfstring{$g$}{g}}\label{sec:constraint-g}

En théorie, le filtre $g$ devrait couper les basses et hautes fréquences car il représente la réponse du microphone, qui est un filtre passe-bande physique.
Cependant, nos études préliminaires montrent que cette contrainte n'était pas respectée au cours des itérations, conduisant parfois à une mauvaise estimation de la RIR.
Pour résoudre ce problème, nous imposons deux zéros, à la fréquence nulle et à la fréquence de Nyquist, dans la fonction de transfert du filtre pour obtenir une meilleure estimation. La matrice $G_{old}$ devient $G=G_1G_0$ avec $G_1=G_{old}$ et $G_0 = \text{Toep}([1, 0, -1])$.

De plus, pour garantir la cohérence du modèle, le filtre $g$ est stabilisé pour maintenir $G^{-1}$ numériquement stable. La version stable de $g$ est obtenue en plaçant tous les pôles de module supérieur à 1 à l'intérieur du cercle unité, c'est à dire en les remplaçant par l'inverse de leur conjugué. Le changement de gain global du filtre n'est pas compensé.

Les paramètres sont initialisés avec $PEG=I_{L_h}$, $R_h=I_{L_h}$, $\mu_h = \delta$, $\sigma_w = 1$, $\sigma_\epsilon = 1$. L'étape de normalisation garantit que $g$ est stable, que $a$ est strictement positif et que $r_h$ est strictement positif pour assurer la cohérence avec le modèle.
Nous résumons l'approche itérative proposée dans l'Algorithme~\ref{alg:algo}.

\begin{algorithm}
    \small
    \caption{Estimation de la RIR proposée}
    \label{alg:algo}
    \begin{algorithmic}[1]
        \STATE \textbf{Initialisation :} initialiser $\theta, z$
        \STATE Nombre d'itérations $I$
        \STATE Optimisateur Adam, tous les paramètres $\theta, z$, $lr=1e-3$
        \FOR{$i = 1$ à $I$}
        \STATE Construire $P, E, G$ en utilisant les paramètres actuels $\theta$
        \STATE Normaliser les paramètres
        \STATE Optimisateur : remise à zéro des gradients
        \STATE Calculer la fonction de coût : $-\mathcal{L}(q,\theta)$
        \STATE Rétropropagation
        \STATE Mettre à jour $\theta, z$
        \ENDFOR
        \STATE \textbf{Retour :} $\theta, z$
    \end{algorithmic}
\end{algorithm}
\section{Configuration expérimentale}

Cette section décrit les données, les métriques et les méthodes utilisées dans notre évaluation. Nous estimons la RIR à partir d'un signal de parole réverbéré, en ajoutant du bruit pour évaluer la robustesse de notre approche.

\subsection{Jeux de données}

\textbf{Jeu de données de parole}
\quad Dans toutes les expériences, nous utilisons une seule source audio, \texttt{87-8000}, sélectionnée aléatoirement dans le jeu de données LibriSpeech \cite{panayotovLibrispeechASRCorpus2015}, tronquée à $2~$s. La fréquence d'échantillonnage originale est de $16~$kHz, mais pour notre évaluation, nous l'avons réduite à $8~$kHz. Des expériences préliminaires menées à $16~$kHz et $8~$kHz ont donné des résultats similaires, nous avons donc opté pour la version $8~$kHz pour des raisons d'efficacité algorithmique.

\noindent\textbf{Jeu de données de RIR}
\quad Pour démontrer la robustesse de notre algorithme, nous utilisons des RIR réelles avec des valeurs de $\text{RT}_{60}$ inférieures à $250$ ms. Les RIR proviennent de la base de données \emph{Aachen Impulse Response}~\cite{jeubBinauralRoomImpulse2009}. Nous avons utilisé 30 RIRs enregistrées dans 5 salles différentes pour 6 configurations de haut-parleurs et de microphones dans chaque cas. Nous considérons 1000 échantillons de RIR, correspondants à $125$ ms. Ce choix est principalement dû au coût de calcul élevé de notre système et parce que nous souhaitions nous concentrer sur l'estimation des premières réflexions, qui est l'aspect le plus difficile de l'estimation de la RIR, plutôt que sur la réverbération tardive.

\noindent\textbf{Jeu de données de bruit}
\quad Comme notre modèle dans l'équation~\eqref{eq:mix} prend en compte un bruit gaussien, nous avons décidé d'ajouter du bruit au signal de parole réverbéré. Le bruit provient du jeu de données WHAMR! \cite{maciejewskiWHAMRNoisyReverberant2020}. Pour chacune des 30 RIRs, un échantillon de bruit de même durée que le signal réverbéré est tiré de WHAMR!, puis on ajuste la puissance du bruit afin d'obtenir le rapport signal à bruit voulu. Nous avons testé plusieurs niveaux de bruit, allant de $20$ dB à $-3$ dB.

\subsection{Métriques et description des tâches}\label{sec:metrics_task}
Nous considérons l'erreur absolue moyenne entre une RIR estimée $\tilde{h}$ et la RIR de référence $h$ pour diverses caractéristiques acoustiques $m \in~$ \{EDC, EDR, RT30\} que nous notons $\Delta_m$ et qui représentent respectivement la courbe de décroissance énergétique, le relief de décroissance énergétique, et le temps de réverbération à $30$ dB. Nous considérons également l'erreur quadratique moyenne (MSE).

\subsection{Méthodes}
Nous testons dans nos expériences l'approche proposée et deux méthodes de référence :
\begin{itemize}
    \item \Our: notre modèle physique variationnel pour la réverbération et l'estimation de la réponse impulsionnelle de salle. Nous regroupons l'initialisation des paramètres et la procédure dans l'Algorithme~\ref{alg:algo}.
    \item \BaselineOne : \textit{déconvolution spectrale} où $y,h,s$ sont considérés dans le domaine temps-fréquence, ce qui donne pour chaque point $f,t$ leurs coefficients de Fourier $H_{f, t}, Y_{f,t}$ et $S_{f,t}$, pour lesquels les opérations suivantes sont calculées : $H_{f, t} = Y_{f, t} / S_{f, t}$.
    \item \BaselineTwo : le filtrage inter-bandes défini dans \cite{avargelSystemIdentificationShortTime2007} avec les paramètres suivants : $K=1$ bande, $n_{fft}=512$, $50$\% de recouvrement et une fenêtre de Hann.
\end{itemize}
\section{Résultats et discussions}

Le \autoref{tab:results} présente les résultats d'erreur pour les 5 configurations de bruit choisies. Pour des raisons de clarté de lecture, l'erreur quadratique moyenne est affichée en pourcentage de la puissance moyenne de la RIR de référence. Il en est de même pour l'écart de temps de réverbération, qui est exprimé en pourcentage du RT$_{30}$ de référence.
Les résultats montrent que \Our\ surpasse les deux méthodes de référence, \BaselineOne\ et \BaselineTwo, pour toutes les métriques et tous les niveaux de SNR inférieurs à $20$ dB.

Ces expériences initiales démontrent que \Our\ a le potentiel d'estimer une RIR et d'extraire des paramètres acoustiques associés. De futures expériences pourraient être réalisées avec des RIR plus longues et des temps de réverbération plus variés pour confirmer ces résultats.

\begin{table}[ht]
    \centering
    \footnotesize
    \caption{\'Ecarts entre la RIR de référence et l'estimation pour les 3 modèles considérés pour 5 SNR différents. Les scores sont présentés sous la forme médiane $\pm $ écart type.}
    \begin{center}
        \begin{tabular}{lccccc}
            \toprule
            Méthode      & $\Delta_{\text{RT}_{30}}\downarrow$ (\%) & $\Delta_{\text{EDC}}\downarrow$    & $\Delta_{\text{EDR}}\downarrow$    & MSE $\downarrow$ (\%)                                       \\
            \midrule
            \Our         & \textbf{89 $\pm$ 50}           & \textbf{0.01 $\pm$ 0.01} & 1.37 $\pm$ 0.23          & \textbf{\hphantom{00}11 $\pm$ \hphantom{000}2} \\
            \BaselineOne & 97 $\pm$ 54                    & 0.09 $\pm$ 0.06          & \textbf{1.03 $\pm$ 0.28} & \hphantom{00}25 $\pm$ \hphantom{00}23          \\
            \BaselineTwo & 96 $\pm$ 53                    & 0.11 $\pm$ 0.06          & 1.09 $\pm$ 0.27          & \hphantom{00}25 $\pm$ \hphantom{00}21          \\
            \multicolumn{5}{c}{{SNR: 20 dB}}                                                                                                                     \\
            \midrule
            \Our         & \textbf{95 $\pm$ 53}           & \textbf{0.03 $\pm$ 0.03} & \textbf{1.50 $\pm$ 0.24} & \textbf{\hphantom{00}14 $\pm$ \hphantom{000}6} \\
            \BaselineOne & 97 $\pm$ 54                    & 0.32 $\pm$ 0.10          & 1.91 $\pm$ 0.48          & \hphantom{0}248 $\pm$ \hphantom{0}229          \\
            \BaselineTwo & 97 $\pm$ 54                    & 0.33 $\pm$ 0.10          & 1.92 $\pm$ 0.49          & \hphantom{0}245 $\pm$ \hphantom{0}206          \\
            \multicolumn{5}{c}{{SNR: 10 dB}}                                                                                                                     \\
            \midrule
            \Our         & \textbf{95 $\pm$ 51}           & \textbf{0.05 $\pm$ 0.03} & \textbf{1.58 $\pm$ 0.36} & \textbf{\hphantom{00}18 $\pm$ \hphantom{000}9} \\
            \BaselineOne & 98 $\pm$ 54                    & 0.44 $\pm$ 0.08          & 2.77 $\pm$ 0.63          & 1245 $\pm$ 1146                                \\
            \BaselineTwo & 98 $\pm$ 54                    & 0.44 $\pm$ 0.07          & 2.76 $\pm$ 0.65          & 1217 $\pm$ 1026                                \\
            \multicolumn{5}{c}{{SNR: 3 dB}}                                                                                                                      \\
            \midrule
            \Our         & \textbf{96 $\pm$ 53}           & \textbf{0.15 $\pm$ 0.10} & \textbf{1.97 $\pm$ 0.38} & \textbf{\hphantom{00}41 $\pm$ \hphantom{00}51} \\
            \BaselineOne & 97 $\pm$ 54                    & 0.45 $\pm$ 0.06          & 3.24 $\pm$ 0.67          & 2223 $\pm$ 2215                                \\
            \BaselineTwo & 96 $\pm$ 55                    & 0.46 $\pm$ 0.07          & 3.24 $\pm$ 0.69          & 2432 $\pm$ 2081                                \\
            \multicolumn{5}{c}{{SNR: 0 dB}}                                                                                                                      \\
            \midrule
            \Our         & 97 $\pm$ 53                    & \textbf{0.22 $\pm$ 0.13} & \textbf{2.30 $\pm$ 0.49} & \textbf{\hphantom{00}71 $\pm$ \hphantom{00}75} \\
            \BaselineOne & 79 $\pm$ 53                    & 0.46 $\pm$ 0.06          & 3.71 $\pm$ 0.74          & 4611 $\pm$ 3712                                \\
            \BaselineTwo & \textbf{65 $\pm$ 52}           & 0.48 $\pm$ 0.06          & 3.76 $\pm$ 0.76          & 4808 $\pm$ 3949                                \\
            \multicolumn{5}{c}{{SNR: -3 dB}}                                                                                                                     \\
            \bottomrule
        \end{tabular}
    \end{center}
    \label{tab:results}
\end{table}
\section{Conclusion et perspectives}

Dans cet article, nous avons proposé une nouvelle méthode pour estimer des réponses impulsionnelles de salles (RIR), dérivée d'un modèle de réverbération stochastique basé sur la physique présenté dans \cite{badeau2018unified}. Notre approche repose sur une formulation variationnelle, qui permet non seulement une estimation précise de la RIR, mais fournit également le filtre correspondant utilisé pour modéliser la RIR. Les résultats obtenus à partir de signaux réverbérés et bruités démontrent l'efficacité de notre estimateur robuste de RIR par rapport aux méthodes de référence traditionnelles.

Dans nos travaux futurs, cette méthode pourra être étendue à des tâches d'adaptation acoustique plus complexes. Le processus de mélange peut être représenté comme un produit de matrices, rendant possible l'application d'une approche variationnelle pour estimer les paramètres pertinents. De plus, nous prévoyons d'essayer d'améliorer la vitesse de convergence, par exemple via un algorithme d'espérance-maximisation où les paramètres sont mis à jour de manière alternée.
Une autre direction pour les travaux futurs consiste à adapter l'approche proposée à la récente théorie ondulatoire statistique, qui établit les propriétés statistiques des solutions de l'équation des ondes dans des domaines bornés \cite{Badeau-24}.

\section*{Remerciements}
Ce travail a été financé par l'ANR SAROUMANE (ANR-22-CE23-0011).

\footnotesize
\bibliography{mybib}


\end{document}